\documentclass[preprint,prd,10pt,aps,showpacs,showkeys]{revtex4-1}
\usepackage[english]{babel}
\usepackage{blindtext}
\usepackage{amsmath,amssymb,bm}
\usepackage{tikz}
\renewcommand{\d}{\mathrm{d}}
\renewcommand{\Re}{\mathrm{Re}\,}
\renewcommand{\Im}{\mathrm{Im}\,}
\newcommand{\e}{\mathrm{e}}

\renewcommand{\i}{\mathrm{i}}
\newcommand{\f}{\varphi}
\newcommand{\Li}{\mathrm{Li}}
\begin{document}

\title{Exact high temperature expansion of the one-loop thermodynamic potential with complex chemical potential}
\date{\today}
\author{Bruno Klajn}
\email{bruno.klajn@irb.hr}
\affiliation{Theoretical Physics Division, Rudjer Bo\v skovi\' c Institute \\ P.O.Box 180, HR-10002 Zagreb, Croatia}
\begin{abstract}
We present a derivation of an exact high temperature expansion for a one-loop thermodynamic potential $\Omega(\tilde{\mu})$ with complex chemical potential $\tilde{\mu}$. The result is given in terms of a single sum the coefficients of which are analytical functions of $\tilde{\mu}$ consisting of polynomials and polygamma functions, decoupled from the physical expansion parameter $\beta m$. The analytic structure of the coefficients permits us to explicitly calculate the thermodynamic potential for the imaginary chemical potential and analytically continue the domain to the complex $\tilde{\mu}$ plane. Furthermore, our representation of $\Omega(\tilde{\mu})$ is particularly well suited for the Landau--Ginzburg-type of phase transition analysis. This fact, along with the possibility of interpreting the imaginary chemical potential as an effective generalized-statistics phase, allows us to investigate the singular origin of the $m^3$ term appearing only in the bosonic thermodynamic potential.
\end{abstract}
\pacs{11.10.Wx, 11.15.Ex, 12.38.Gc}
\keywords{finite-temperature field theory, partition function, thermodynamic potential, high temperature expansion, imaginary chemical potential, generalized statistics, generalized boundary conditions, phase transition}
\maketitle
\section*{Introduction}
Quantum field theory at finite temperature is known to be very successful at dealing with problems such as phase transitions, spontaneous symmetry breaking and restoration, and similar collective phenomena in the context of high energy physics \cite{kapusta,Bellac}. In such applications, a quantity of key interest is the thermodynamic potential which to one-loop order is proportional to
\begin{equation}
\Omega_{\text{B/F}} = \frac{2}{\beta} \int{\frac{\d^3 p}{(2\pi)^3} \ln\left(1 \mp \e^{- \beta E}\right)},
\label{TheIntegral0}
\end{equation}
where $\beta > 0$ is the inverse temperature and $E=\sqrt{\bm{p}^2+m^2}$ is the relativistic energy of the field of mass $m>0$. Here, the upper (lower) sign corresponds to Bose (Fermi) statistics. It should be noted that the vacuum contribution to $\Omega_{\text{B/F}}$ is neglected since it is temperature independent and as such is irrelevant for our considerations. The thermodynamic potential can be calculated approximately in the limit of low $(\beta m \gg 1)$ or high temperature $(\beta m \ll 1)$. In our notation, the original result for $\Omega_{\text{B/F}}$ in the high temperature limit, given by Dolan and Jackiw \cite{Dolan:1973qd}, reads
\begin{equation}
\Omega_{\text{B}} \approx -\frac{\pi^2}{45 \beta^4} + \frac{1}{12\beta^2}m^2 - \frac{1}{6\pi\beta}m^3 + \frac{1}{32\pi^2} \left(\frac{3}{2} - 2\gamma + 2 \ln 4\pi - \ln \beta^2 m^2\right)m^4,
\label{Bos}
\end{equation}
for bosons and
\begin{equation}
\Omega_{\text{F}} \approx \frac{7\pi^2}{360\beta^4} - \frac{1}{24\beta^2}m^2 + \frac{1}{32\pi^2}\left(\frac{3}{2}-2\gamma+2\ln \pi - \ln \beta^2 m^2 \right)m^4,
\label{Fer}
\end{equation}
for fermions. Here, $\gamma$ is the Euler-Mascheroni constant. One of the most striking differences in the above results is the appearance of $m^3$ term in $\Omega_{\text{B}}$, the expression otherwise being very similar to $\Omega_{\text{F}}$. It is evident that the origin of the term in question is `singular' since the original integral is dependent only on $m^2$ (through $E$), and all other terms in the high temperature expansion are indeed functions of $m^2$. This is reminiscent of the Bose-Einstein condensation (cf.~\cite{Haber:1981ts,Haber:1981tr}) which is also, in a certain sense, a singular phenomenon, occurring only in bosonic systems. Ultimately, the $m^3 \equiv (m^2)^{3/2}$ term is a manifestation of a branch cut in $\Omega_{\text{B}}$ in the variable $m^2$  running from $m^2 = 0$ to $m^2 = - \infty$. Such a branch cut does not appear in $\Omega_{\text{F}}$. Given the similarities and the differences in the expressions for $\Omega_{\text{B}}$ and $\Omega_{\text{F}}$, it is natural to ask, is there a way of calculating both potentials at once?

The above discussion motivates us to consider the thermodynamic potential given in Eq.~\eqref{TheIntegral0} as a special case of a generalized thermodynamic potential
\begin{equation}
\Omega(\f) = \frac{1}{\beta} \int{\frac{\d^3 p}{(2\pi)^3} \ln\left(1-\e^{\i \f} \e^{- \beta E}\right)} + \mathrm{c.c.},
\label{TheIntegral}
\end{equation}
where $\f \in [0,2\pi \rangle$ is an effective generalized-statistics parameter which interpolates between bosons ($\varphi=0$) and fermions ($\varphi=\pi$). The c.c.~stands for complex conjugation and corresponds to the antiparticle contribution to $\Omega(\f)$. Since we are dealing with thermodynamic potential in (3+1) dimensions, the term generalized statistics should not be confused with the anyon statistics which is only relevant in (2+1) dimensions. The parameter $\f$ is directly related to the imaginary chemical potential $\mu_\i$ ($\mu_{\i} = \f/\beta$ for bosons and $\mu_{\i} = (\f+\pi)/\beta$ for fermions), which is especially relevant for finite temperature quantum chromodynamics (QCD). The imaginary chemical potential can be used to obtain the information about the QCD phase diagram and is particularly well suited for the lattice calculations \cite{Roberge:1986mm,Hart:2000ef,deForcrand:2002ci,deForcrand:2010he,Aarts:2013bla}. Furthermore, the Polyakov loop is the order parameter in deconfinement phase transitions, where it plays the role of the imaginary chemical potential \cite{Polyakov:1978vu,Weiss:1980rj,Weiss:1981ev,Fukushima:2003fw,Ratti:2005jh} in the mean field approximation. The imaginary chemical potential is also used to calculate the so-called dressed Polyakov loop \cite{Gattringer:2006ci,Synatschke:2007bz,Bilgici:2008qy} which is used, in particular, to study the connection between the chiral and deconfinement phase transitions \cite{Fischer:2009gk,Fischer:2009wc,Kashiwa:2009ki,Mukherjee:2010cp,Gatto:2010qs,Benic:2013zaa}.

Integrals of the type \eqref{TheIntegral0} and \eqref{TheIntegral} have been investigated by several authors (cf. \cite{Braden:1981we,Actor:1985zf,Weldon:1985yh,Actor:1987cf,Meisinger:2001fi} and references therein) in the past. The first step in all the calculations is performing the angular integration, followed by a change of integration variable and then a partial integration. This transforms Eq.~\eqref{TheIntegral} into
\begin{equation}
\Omega(\f) = -\frac{m^4}{6\pi^2}\int_1^\infty{\frac{(t^2-1)^{3/2}}{\e^{-\i\varphi} \e^{\beta m t}-1} \d t} + \mathrm{c.c.}.
\label{Fifth}
\end{equation}
From the form of the Eq.~\eqref{Fifth}, it is clear that if one allows $\f$ to be a complex number $\f = \f_\text{r} + \i \f_\text{i}$ with $\f_\text{r} \in [0,2\pi \rangle$ and $\f_\text{i} \in \mathbb{R}$, the function $\Omega(\f)$ has two branch cuts in the restriction of the complex plane $\mathbb{C}' = [0,2\pi \rangle \times \i \mathbb{R}$ located on the imaginary axis and running from $\pm \i \beta m$ to $\pm \i \infty$. Therefore, for non-zero values of $\beta m$, the potential $\Omega(\f)$ is analytic in the neighborhood of $\f = 0$ and an analytic continuation from real $\f$ to complex $\f$ can be performed.

Next, taking into account that $|\e^{\i\varphi} \e^{-\beta m t}|<1$, the denominator is expanded in (a uniformly convergent) Taylor series and a termwise integration is performed which, upon using the integral representation of the modified Bessel function of second kind
\begin{equation}
K_n(x) = \frac{\sqrt{\pi}}{\Gamma(n+\frac{1}{2})} \left(\frac{x}{2}\right)^n \int_1^\infty \d t\, \e^{-t x} (t^2-1)^{n-1/2},
\label{Bessel}
\end{equation}
turns Eq.~\eqref{TheIntegral} into a sum over Bessel functions
\begin{equation}
\Omega(\f) = -\frac{m^2}{2\pi^2 \beta^2} \sum_{n=1}^{\infty}{\frac{\e^{\i n \varphi}}{n^2}} K_2(n \beta m) + \mathrm{c.c.}.
\label{TheSum}
\end{equation}

There are three different approaches, known to the author, which make it possible to transform the sum of Bessel functions into a more tractable result. It should be noted, however, that the original approximate results for $\Omega_{\text{B/F}}$, given by Eqs.~\eqref{Bos} and \eqref{Fer}, were obtained directly from Eq.~\eqref{TheIntegral0}, without the use of Bessel functions. In the first approach, an exact series representation of the integral $\Omega(\f)$ is limited to boson and fermion statistics and given in Ref.~\cite{Braden:1981we} using Mellin transform. The result obtained is of the form
\begin{equation}
\Omega_{\text{B/F}} = \sum_k (a_{\text{B/F}})_k (\beta m)^k,
\label{eq:Braden}
\end{equation}
and correctly reproduces the original approximate results \cite{Dolan:1973qd}. The second approach consists in substituting the series representation of the Bessel function into the Eq.~\eqref{TheSum} and performing the termwise sum over $n$ making the use of the zeta regularization techniques. This method was first employed, in the special case of bosons and fermions, in Refs.~\cite{Actor:1985zf,Weldon:1985yh}. The final result is given in terms of an expansion in the parameter $\beta m$, in agreement with Ref.~\cite{Braden:1981we}. The method was later generalized in Ref.~\cite{Actor:1987cf} for the case of arbitrary parameter $\f$ by expanding the exponential term $\e^{\i n \f}$ of Eq.~\eqref{TheSum} in Taylor series. This led to a double series expansion in the parameters $\beta m$ and $\f$,
\begin{equation}
\Omega(\f) = \sum_{k,m} a_{km} (\beta m)^k \f^m.
\label{eq:Actor}
\end{equation}
The author of Ref.~\cite{Actor:1987cf}, in fact, has also considered the non-zero real chemical potential $\mu$ and obtained the triple series expansion in parameters $\beta m$, $\f$ and $\mu$. Finally, the third method of resumming Eq.~\eqref{TheSum} is given in Ref.~\cite{Meisinger:2001fi}. It utilizes various Bessel function identities to convert the sum over Bessel functions to a sum over elementary functions (integer powers and square roots) containing both the $\beta m$ and $\f$ parameters combined in a nontrivial way as
\begin{equation}
\Omega(\f) = \sum_k f_k (\beta m, \f).
\label{eq:Meisinger}
\end{equation}

Although Eqs.~\eqref{eq:Actor} and \eqref{eq:Meisinger} generalize the result \eqref{eq:Braden} for arbitrary $\f$, it is of special interest to express generalized result in a power expansion of the form
\begin{equation}
\Omega(\f) = \sum_k a_k(\f) (\beta m)^k,
\label{TheForm}
\end{equation}
with the coefficients $a_k(\f)$ being simple analytic functions of the phase $\f$. Namely, the main benefits of such a form are: it is a single sum expression, which is an advantage in comparison to form \eqref{eq:Actor} and the parameter $\f$ is `decoupled' from the physical variable $\beta m$ which is physically more operational and transparent than the form \eqref{eq:Meisinger}. The form in Eq.~\eqref{TheForm} is particularly suitable for Landau--Ginzburg-type analysis of phase transitions for arbitrary $\f$. Moreover, in electroweak baryogenesis, the appearance of the $m^3$ term in Eq.~\eqref{Bos} is necessary for the electroweak phase transition to be of the first order (see e.g. Ref.~\cite{Funakubo:1996iw} and references therein). Therefore, it is reasonable to expect that Eq.~\eqref{TheForm} will produce a deeper insight on the origin and properties of the $m^3$ term in the expression for $\Omega_{\text{B}}$.

In this paper, we derive the exact series representation of the thermodynamic potential $\Omega(\f)$ of the type given in Eq.~$\eqref{TheForm}$. The final expression is given in Eq.~\eqref{TheResult}, where the coefficients $a_k(\f)$ are given in terms of polynomials and polygamma functions. With expression \eqref{TheResult} in hand, we are in a position to show that the singular $m^3$ term is absent in $\Omega(\f)$ for all $\f \neq 0$ but appears naturally in the limit $\f \to 0$. Also, due to a simple analytic structure of $a_k(\f)$, we generalize our results to the case of complex chemical potential $\tilde{\mu} = \mu + \i \f/\beta$ in Eq.~\eqref{FinalResult}, thereby obtaining for the first time the complete dependence of the thermodynamic potential $\Omega(\tilde{\mu})$ on the complex chemical potential $\tilde{\mu}$ which is the main result of this paper.

The rest of the paper is organized as follows: in Section I, we obtain the exact high temperature expansion of the thermodynamic potential $\Omega(\f)$ for $\f \neq 0$ using a generalization of the zeta-regularization method of Refs.~\cite{Actor:1985zf,Weldon:1985yh}. In Section II, we explicitly show that $\lim_{\f \to 0} \Omega(\f) = \Omega_{\text{B}}$, and that the singular term in $\Omega_{\text{B}}$ containing $m^3$ is generated by a resummation of a divergent series generated by the coefficients $a_k(\f)$ in the limit $\f \to 0$. Finally, in Section III, we use the analyticity of $\Omega(\f)$ to extend the domain of the thermodynamic potential to the complex $\f$ plane $\Omega(\f) \to \Omega(\tilde{\mu})$, with $\tilde{\mu} = \mu + \i \f/\beta$, as a result of which it becomes a function of the real $\mu$, as well as the imaginary chemical potential $\mu_\i$. At the end, we discuss the relation between our result and the results obtained earlier and give an outlook to possible future work. Some calculation details are given in an Appendix.

\section{Calculating the thermodynamic potential $\Omega(\f)$ for $\f \neq 0$}

\subsection{Expanding the modified Bessel function}

To calculate the thermodynamic potential given in Eq.~\eqref{TheSum}, we use the series representation of the modified Bessel function valid for $x>0$ \cite{abramowitz1970handbook},
\begin{equation}
K_2(x) = \frac{2}{x^2}-\frac{1}{2} +\frac{1}{2}\sum_{k=0}^\infty \frac{\psi(k+1)+\psi(k+3)- 2 \ln \frac{x}{2}}{\Gamma(k+1)\Gamma(k+3)}\left(\frac{x^2}{4}\right)^{k+1},
\end{equation}
where the digamma function $\psi(z) = \frac{\d}{\d z} \ln \Gamma(z)$ is the logarithmic derivative of the gamma function. Plugging this expression in Eq.~\eqref{TheSum}, we obtain
\begin{equation}
\Omega(\f) = -\frac{m^2}{2\pi^2 \beta^2} \left\{\frac{2}{\beta^2 m^2}\sum_{n=1}^\infty {\frac{\e^{\i n \f}}{n^4}} - \frac{1}{2} \sum_{n=1}^\infty {\frac{\e^{\i n \f}}{n^2}} + \frac{1}{2} \sum_{n=1}^\infty {\e^{\i n \f}} \sum_{k=0}^\infty \left[A(k)-2B(k) \ln n \right]n^{2k} \right\} + \mathrm{c.c.},
\label{Expanded}
\end{equation}
where we have defined two auxiliary functions,
\begin{equation}
A(k) = \frac{\psi(k+1)+\psi(k+3)- \ln \frac{\beta^2 m^2}{4}}{\Gamma(k+1)\Gamma(k+3)}\left(\frac{\beta^2 m^2}{4}\right)^{k+1},
\end{equation}
and
\begin{equation}
B(k) = \frac{1}{\Gamma(k+1)\Gamma(k+3)}\left(\frac{\beta^2 m^2}{4}\right)^{k+1}.
\end{equation}
Due to the nature of the gamma and digamma functions, these functions are entire in the complex $k$ plane.

The two single sums inside the curly bracket in Eq.~\eqref{Expanded} are convergent and can be immediately identified as polylogarithms \cite{abramowitz1970handbook,lewin1981polylogarithms}
\begin{equation}
\sum_{n=1}^\infty {\frac{\e^{\i n \f}}{n^4}} = \Li_4(\e^{\i \f}), \quad \sum_{n=1}^\infty {\frac{\e^{\i n \f}}{n^2}} = \Li_2(\e^{\i \f}).
\label{eq:Convergent}
\end{equation}
To perform the sum over $n$ in the double sum terms in Eq.~\eqref{Expanded}, one would first have to interchange the order of summation to find that the sum over $n$ diverges. Therefore, the naive summation swap does not work. To correctly sum over $n$, we follow (and generalize) the method of Refs.~\cite{Weldon:1985yh,Actor:1987cf,Elizalde:1988xc}.

\subsection{Interchanging the order of summations}

\begin{figure}[!tb]
\centering
\includegraphics{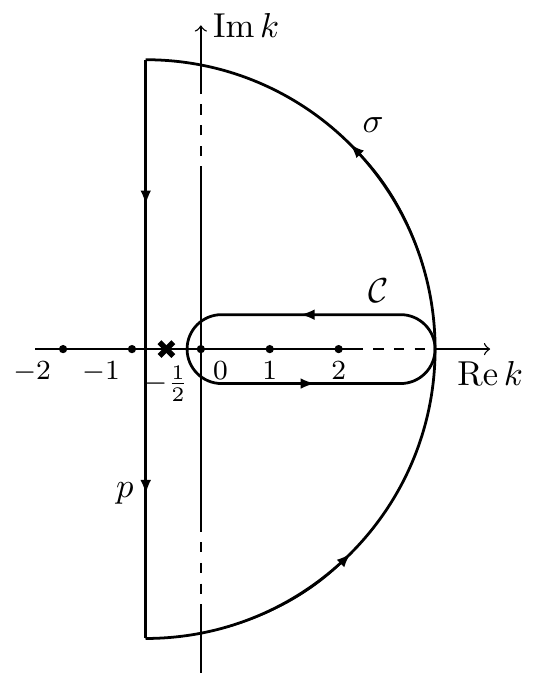}
\caption{The contours of integration in the complex $k$ plane. Dots represent the poles of the cotangent function, while \textbf{x} represents the summation-induced pole of the polylogaritm functions for $\f = 0$.}
\label{Figure}
\end{figure}

Due to the fact that the functions $A(k)$ and $B(k)$ are regular in the complex $k$ plane, the residue theorem can be used to write the sum over $k$ in Eq.~\eqref{Expanded} as an contour integral
\begin{equation}
\sum_{k=0}^\infty \left[A(k)-2B(k) \ln n \right]n^{2k} = \frac{1}{2\i} \oint_{\mathcal{C}}{\d k \left[A(k)-2B(k) \ln n \right]n^{2k} \cot \pi k},
\end{equation}
where the curve $\mathcal{C}$ encompasses the positive real axis $\Re k$ (a Hankel curve). The poles of the cotangent function inside the curve reproduce the original sum over $k$. Again, due to the regularity of the integrand, the curve $\mathcal{C}$ can be deformed into a perimeter of a half-disk consisting of a straight line $p$, defined by $\Re k = - \epsilon$, $\epsilon \in \langle 0, 1 \rangle$, and a semicircle $\sigma$ in the right half-plane (see Fig.~\ref{Figure}.). The integral over $\sigma$ vanishes \cite{Actor:1987cf}, and only the integral over $p$ contributes.

To interchange the sum over $n$ with the integral over $k$, the sum must converge. This is the case for $\Re (2k) < -1$, i.e. for $\epsilon \in \langle\frac{1}{2},1\rangle$. Therefore, if the line $p$ is at a distance of at least $\frac{1}{2}$ from the origin of the complex $k$ plane, the sum over $n$ can be safely introduced into the integral over $k$, with the result
\begin{equation}
\sum_{n=1}^\infty {\frac{\e^{\i n \f}}{n^{-2k}}} = \Li_{-2k}(\e^{\i \f}),
\end{equation}
and
\begin{equation}
\sum_{n=1}^\infty {\frac{\e^{\i n \f}}{n^{-2k}} \ln n} = - \frac{\partial}{\partial s} \sum_{n=1}^\infty {\frac{\e^{\i n \f}}{n^s}}\Big|_{s=-2k} \equiv - \Li'_{-2k}(\e^{\i \f}).
\label{derpol}
\end{equation}

On the line $p$, all sums over $n$ produce non-positive integer order polylogarithms and their derivatives with respect to the order. Now, we simply extend the validity of this summation by analytical continuation over the whole complex plane $k$. This is, in effect, a generalization of the zeta function regularization \cite{Elizalde:1994gf}, since the polylogarithm is a generalization of the Riemann zeta function, in such a way that 
\begin{equation}
\Li_s(1) = \zeta(s).
\end{equation}
Furthermore, the polylogarithm of a variable defined on the unit complex circle, as in this case, is sometimes called the periodic zeta function \cite{apostol1990modular}
\begin{equation}
F(\f/2\pi,s) = \Li_s(\e^{\i \f}).
\end{equation}
This function is regular as a function of $s$ for fixed $\f \neq 0$, and has a simple pole at $s=1$ for $\f = 0$, when it simplifies to the ordinary zeta function. This means that in the purely boson case, the introduction of the sum over $n$ under the integral over $k$ produces an additional simple pole at $k = -\frac{1}{2}$ (see Fig.~\ref{Figure}.).

After the summation over $n$ is performed, the integral over the semicircle $\sigma$ again vanishes (including the `worst-case-scenario' $\f = 0$ \cite{Elizalde:1988xc,Elizalde:1994gf}), so it can be safely added to the integral over $p$ to form a closed contour. Let us, for the time being, assume $\f \neq 0$, so that we can again deform the contour to the original Hankel contour $\mathcal{C}$ and finally obtain
\begin{equation}
\Omega(\f)\big|_{\f \neq 0} = -\frac{m^2}{2\pi^2 \beta^2} \Bigg\{\frac{2}{\beta^2 m^2} \Li_4(\e^{\i \f}) - \frac{1}{2} \Li_2(\e^{\i \f}) + \frac{1}{2} \sum_{k=0}^\infty \left[A(k) \Li_{-2k}(\e^{\i \f}) + 2 B(k) \Li'_{-2k}(\e^{\i \f}) \right] \Bigg\} + \mathrm{c.c.}.
\label{Interchanged}
\end{equation}

Equation \eqref{Interchanged} shows that for $\f \neq 0$ we could have simply interchanged the order of $k$ and $n$ sums and regularize the sum over $n$ using the polylogarithms. However, this procedure would not work for $\f = 0$ (this was first shown in Ref.~\cite{Weldon:1985yh}), as we will see later in greater detail. We next concentrate on simplifying the expressions containing polylogarithms and their derivatives.

\subsection{Simplifying the polylogarithms}

Since the polylogarithms (together with their derivatives) in Eq.~\eqref{Interchanged} appear in complex conjugate pairs, they can be simplified using the relation between the polylogarithms and the Hurwitz zeta function $\zeta(s,x)$ \cite{abramowitz1970handbook,lewin1981polylogarithms}
\begin{equation}
\i^{-s} \Li_s(\e^{\i\f}) + \i^s \Li_s(\e^{-\i\f}) = \frac{(2\pi)^s}{\Gamma(s)}\zeta(1-s,\f/2\pi),
\label{Hurwitz}
\end{equation}
which holds for $s \in \mathbb{C}$, $\Re \f \in [0,2\pi\rangle$, $\Im \f \geq 0$. In the case $\Im \f < 0$, for the relation to be valid, it is necessary that $\Re \f \in \langle0,2\pi]$. For $s = n \in \mathbb{N}_0$, the Hurwitz zeta function is related to the Bernoulli polynomials \cite{abramowitz1970handbook} via
\begin{equation}
\zeta(-n,x) = - \frac{B_{n+1}(x)}{n+1}.
\label{Bernoulli}
\end{equation}
Equations \eqref{Hurwitz} and \eqref{Bernoulli} can be directly used for $s=2,4$ while for $s=0$ we find
\begin{equation}
\Li_0(\e^{\i \f}) + \Li_0(\e^{-\i \f}) = \lim_{s \to 0} \frac{(2\pi)^s}{\Gamma(s)}\zeta(1-s,\f/2\pi) = -1.
\label{limit}
\end{equation}
For $s = -2k, \ k \in \mathbb{N}$, the sum of polylogarithms vanishes due to the poles of the gamma function. This completes the simplification of the polylogarithms.

As for the derivatives of polylogarithms, we differentiate Eq.~\eqref{Hurwitz} with respect to $s$ and find
\begin{equation}
\label{Derivative}
\Li'_{-2k}(\e^{\i \f}) + \Li'_{-2k}(\e^{-\i \f}) = \frac{\i \pi}{2}\left(\Li_{-2k}(\e^{\i \f}) - \Li_{-2k}(\e^{-\i \f})\right) + (-1)^k \frac{\partial}{\partial s} \left[\frac{(2\pi)^s}{\Gamma(s)}\zeta(1-s,\f/2\pi)\right]_{s=-2k},
\end{equation}
where $k \in \mathbb{N}_0$. The first term on the r.h.s.~can be calculated using the recurrence relation for the polylogarithms
\begin{equation}
\Li_{-n}(z) = \left(z\frac{\partial}{\partial z}\right)^n \Li_{0}(z) \equiv \left(z\frac{\partial}{\partial z}\right)^n \frac{z}{1-z},
\end{equation}
which in our case leads to
\begin{equation}
\Li_{-2k}(\e^{\i \f}) = -\frac{1}{2}\delta_{k0} + (-1)^k \frac{\i}{2} \left(\frac{\partial}{\partial \f}\right)^{2k} \cot \frac{\f}{2}.
\end{equation}
The unpleasant $2k$-th derivative of the cotangent can be dealt with by using the relation
\begin{equation}
\pi \left(\frac{\partial}{\partial z}\right)^n \cot \pi z = (-1)^n \psi^{(n)}(1-z) - \psi^{(n)}(z),
\end{equation}
which can be derived by the repeated differentiation of the reflection formula for the gamma function. Here $\psi^{(n)}(z)$ is the polygamma function, the $(n+1)$-th logarithmic derivative of the gamma function and we have $\psi^{(0)}(z) \equiv \psi(z)$. As a result, the difference of polylogarithms can be written as
\begin{equation}
\Li_{-2k}(\e^{\i \f}) - \Li_{-2k}(\e^{-\i \f}) =  \frac{\i}{\pi} \frac{(-1)^k}{(2\pi)^{2k}}\left[\psi^{(2k)}(1-\f/2\pi) - \psi^{(2k)}(\f/2\pi)\right].
\end{equation}

For the second term on the r.h.s.~of Eq.~\eqref{Derivative}, the cases $k=0$ and $k>0$ should be treated separably. For $k = 0$, we expand the function around $s = 0$ to obtain
\begin{equation}
\frac{\partial}{\partial s} \left[\frac{(2\pi)^s}{\Gamma(s)}\zeta(1-s,\f/2\pi)\right]_{s=0} = - \frac{\partial}{\partial s} \left[(1 + s \ln 2\pi) (s + \gamma s^2) \left(\frac{1}{s} + \psi (\f/2\pi)\right)\right]_{s=0} = -(\ln 2\pi + \gamma + \psi(\f / 2\pi)).
\end{equation}
On the other hand, for $k > 0$, we can perform the differentiation and note that due to the gamma function in the denominator, most of the terms vanish, so that
\begin{equation}
\frac{\partial}{\partial s} \left[\frac{(2\pi)^s}{\Gamma(s)}\zeta(1-s,\f/2\pi)\right]_{s=-2k} = - \frac{1}{(2\pi)^{2k}}\zeta(2k+1,\f/2\pi)\frac{\psi(s)}{\Gamma(s)}\Bigg|_{s=-2k} = \frac{(2k)!}{(2\pi)^{2k}}\zeta(2k+1,\f/2\pi).
\end{equation}
Therefore, the second term in the r.h.s.~in Eq.~\eqref{Derivative} has also been resolved.

As a final touch of simplification, we utilize the relation between the polygamma function and Hurwitz zeta function for integer $k>0$, which is
\begin{equation}
\zeta(2k+1,\f/2\pi) = -\frac{1}{(2k)!} \psi^{(2k)}(\f/2\pi),
\end{equation}
to finally obtain
\begin{equation}
\Li'_{-2k}(\e^{\i \f}) + \Li'_{-2k}(\e^{-\i \f}) = - \delta_{k0} (\gamma + \ln 2\pi) - \frac{1}{2} \frac{(-1)^k}{(2\pi)^{2k}}\left[\psi^{(2k)}(\f/2\pi) + \psi^{(2k)}(1-\f/2\pi)\right].
\label{finob}
\end{equation}

\subsection{The general result for $\Omega(\f)\big|_{\f \neq 0}$}

Putting together Eqs.~\eqref{Expanded}, \eqref{eq:Convergent}, \eqref{derpol}, \eqref{Hurwitz}, \eqref{Bernoulli}, \eqref{limit} and \eqref{finob}, we arrive at a simple series representation for the thermodynamic potential $\Omega$ in the generic $\f \neq 0$ case,
\begin{align}
\Omega(\f)\big|_{\f \neq 0} &= \frac{2\pi^2}{3\beta^4}B_4(\f/2\pi) + \frac{1}{2\beta^2}B_2(\f/2\pi)m^2 + \frac{1}{32\pi^2}\left(\frac{3}{2}+2\ln 4\pi - \ln \beta^2 m^2 \right)m^4 \nonumber \\
&+ \frac{m^4}{16 \pi^2} \sum_{k=0}^\infty \frac{(-1)^k}{k!(k+2)!} \left[\psi^{(2k)}(\f/2\pi) + \psi^{(2k)}(1-\f/2\pi)\right]\left(\frac{\beta m}{4\pi}\right)^{2k},
\label{TheResult}
\end{align}
where we have used
\begin{equation}
A(0) = \frac{\beta^2 m^2}{8}\left(\frac{3}{2}-2\gamma - \ln \frac{\beta^2 m^2}{4}\right).
\end{equation}

This result has several interesting features which are worth pointing out. First of all, it is an exact high temperature series representation of the thermodynamic potential given by \eqref{TheIntegral}. Second, the parameter $\f$ appears only in the coefficients of the series and is, therefore, decoupled from the physical parameter $\beta m$. Third, due to the symmetry of the Bernoulli polynomials, $B_n(1/2 + x) = B_n(1/2 - x)$, we have the (expected) symmetry property for $\Omega(\f)$, namely
\begin{equation}
\Omega(\f) = \Omega(2\pi - \f).
\label{TheSymmetry}
\end{equation}

The most relevant special case of Eq.~\eqref{TheResult} is the fermion thermodynamic potential, for which $\f/2\pi = 1/2$. For this value of $\f$, the Bernoulli polynomials take on the values $B_4(1/2) = \frac{7}{240}$ and $B_2(1/2) = -\frac{1}{12}$, while the polygamma function can be written in terms of the Riemann zeta function
\begin{equation}
\psi^{2k}(1/2)=-\delta_{k0}(\gamma + 2\ln 2) - \bar{\delta}_{k0}(2k)!(2^{2k+1}-1)\zeta(2k+1),
\end{equation}
where we have introduced the notation $\bar{\delta}_{ij} = 1-\delta_{ij}$. This amounts to the following result
\begin{align}
\Omega_{\text{F}} &= \frac{7\pi^2}{360\beta^4} - \frac{1}{24\beta^2}m^2 + \frac{1}{32\pi^2}\left(\frac{3}{2}-2\gamma+2\ln \pi - \ln \beta^2 m^2 \right)m^4 \nonumber \\
&- \frac{m^4}{8 \pi^2} \sum_{k=1}^\infty \frac{(-1)^k}{k!} \frac{(2k)!}{(k+2)!}(2^{2k+1}-1)\zeta(2k+1) \left(\frac{\beta m}{4\pi}\right)^{2k},
\label{TheFermion}
\end{align}
in agreement with the previously obtained results \cite{Dolan:1973qd,Braden:1981we,Weldon:1985yh,Actor:1987cf,Meisinger:2001fi}. It should be noted that the above series has a finite radius of convergence, namely, it converges for $\beta m < \pi$.

While the result of Eq.~\eqref{TheResult} covers almost all of the domain of the parameter $\f$, it leaves out the crucial point $\f = 0$, which, however, is of key importance in physics, corresponding to the case of the boson gas. In the next section, we examine this particular value of $\f$ in great detail.

\section{Calculating the thermodynamic potential $\Omega(\f)$ for $\f = 0$}

\subsection{A direct calculation}

To calculate the thermodynamic potential $\Omega(\f)$ for $\f = 0$, we start with Eq.~\eqref{Interchanged}. As stated earlier, in the boson case, all the polylogarithms reduce to the Riemann zeta function and the contour $\mathcal{C}$ picks up a pole at $k = -\frac{1}{2}$. Denoting by $\Delta$ the contribution of that pole, we have
\begin{equation}
\Omega_{\text{B}} \equiv \Omega(\f)\big|_{\f = 0} = -\frac{m^2}{2\pi^2 \beta^2} \Bigg\{\frac{2}{\beta^2 m^2} \zeta(4) - \frac{1}{2} \zeta(2) + \frac{1}{2} \Delta + \frac{1}{2} \sum_{k=0}^\infty \left[A(k) \zeta(-2k) + 2 B(k) \zeta'(-2k) \right] \Bigg\} + \mathrm{c.c.},
\label{Additional}
\end{equation}
where
\begin{equation}
\Delta = 2\pi\i \, \mathrm{Res}\, \left[\frac{1}{2\i} (A(k) \zeta(-2k) + 2B(k) \zeta'(-2k))\cot \pi k\right]_{k=-\frac{1}{2}} = \frac{\pi^2}{2} B(-1/2) = \frac{\pi}{3} \beta m.
\label{Delta}
\end{equation}
Using the known values of the zeta function 
\begin{equation}
\zeta(4) = \frac{\pi^4}{90},\ \ \zeta(2) = \frac{\pi^2}{6},\ \ \zeta(-2k) = -\frac{1}{2} \delta_{k0},
\end{equation}
and expressing its derivatives as
\begin{equation}
\zeta'(-2k) = - \delta_{k0} \frac{1}{2} \ln 2\pi  + \bar{\delta}_{k0} \frac{(-1)^k}{2} \frac{(2k)!}{(2\pi)^{2k}} \zeta(2k+1),
\end{equation}
the Eq.~\eqref{Additional} takes the form
\begin{align}
\Omega_{\text{B}} &= -\frac{\pi^2}{45 \beta^4} + \frac{1}{12\beta^2}m^2 - \frac{1}{6\pi\beta}m^3 + \frac{1}{32\pi^2} \left(\frac{3}{2} - 2\gamma + 2 \ln 4\pi - \ln \beta^2 m^2\right)m^4 \nonumber \\
&- \frac{m^4}{8\pi^2}\sum_{k=1}^\infty \frac{(-1)^k}{k!} \frac{(2k)!}{(k+2)!} \zeta(2k+1) \left(\frac{\beta m}{4\pi}\right)^{2k}.
\label{TheBoson}
\end{align}
This result, as well, confirms the earlier calculations in Refs.~\cite{Dolan:1973qd,Braden:1981we,Weldon:1985yh,Actor:1987cf,Meisinger:2001fi}. Similarly to the fermion potential $\Omega_{\text{F}}$ of Eq.~\eqref{TheFermion}, the series representation of $\Omega_{\text{B}}$ converges only for $\beta m < 2\pi$.

Equation \eqref{TheBoson} raises an interesting question. Given the above calculations, it appears that the boson statistics, $\f = 0$, is very distinct in comparison with the generic phase $\f \neq 0$ of which the fermion statistics is a special instance. As pointed out before (and now becoming even more apparent), the $m^3$ term in the series expansion for $\Omega(\f)$ is present only in the boson case, $\f = 0$. Is it the case that almost all generalized statistics are fermion-like, and bosons are some kind of an exception? The following calculation will show that the answer is -- no.

\subsection{Deriving $\Omega_{\text{B}}$ from $\Omega(\f)\big|_{\f \neq 0}$}

If the bosons were an exceptional statistics, we would expect the thermodynamic potential $\Omega(\f)$ not to be continuous at $\f = 0$. Indeed, given that the $m^3$ term appears only for $\f = 0$, we have every reason to believe this to be the case. However, the original integral in Eq.~\eqref{TheIntegral} does not appear problematic around $\f = 0$. The only way to resolve this dilemma is by examining the limit $\f \to 0$ of Eq.~\eqref{TheResult}.

In this limit, the Bernoulli polynomials are finite and yield $B_4(0) = -\frac{1}{30}$ and $B_4(0) = \frac{1}{6}$, while the polygamma functions are unbounded
\begin{equation}
\psi^{(2k)}(\f/2\pi) + \psi^{(2k)}(1-\f/2\pi) \approx -(2k)! \left(\frac{2\pi}{\f}\right)^{2k+1} - 2\left(\delta_{k0} \gamma + \bar{\delta}_{k0} (2\pi)! \zeta(2k+1)\right) + \mathcal{O}(\f), \quad \f \to 0.
\end{equation}
Plugging these limits in Eq.~\eqref{TheResult}, we find
\begin{align}
\lim_{\f \to 0} \Omega(\f) &= -\frac{\pi^2}{45\beta^4} + \frac{1}{12\beta^2} + \frac{1}{32\pi^2}\left(\frac{3}{2}-2\gamma +2\ln 4\pi - \ln \beta^2 m^2 \right)m^4 \nonumber \\
&- \frac{m^4}{8 \pi^2} \sum_{k=1}^\infty \frac{(-1)^k}{k!} \frac{(2k)!}{(k+2)!} \zeta(2k+1) \left(\frac{\beta m}{4\pi}\right)^{2k} \nonumber \\
&- \lim_{\f \to 0} \frac{m^4}{8\pi \f} \sum_{k=0}^\infty \frac{(-1)^k}{k!} \frac{(2k)!}{(k+2)!} \left(\frac{\beta m}{2 \f}\right)^{2k}.
\label{TheLimit}
\end{align}
The sum under the limit is evaluated in the Appendix, and is given by 
\begin{equation}
\sum_{k=0}^\infty \frac{(-1)^k}{k!} \frac{(2k)!}{(k+2)!} \left(\frac{\beta m}{2 \f}\right)^{2k} \approx \frac{4\f}{3\beta m}, \quad \f \to 0.
\label{Approx}
\end{equation}
Consequently, we arrive to an important conclusion that the thermodynamic potential is indeed continuous at $\f = 0$, i.e.,
\begin{equation}
\lim_{\f \to 0} \Omega(\f) = \Omega_{\text{B}},
\label{Continuous}
\end{equation}
which is consistent with the fact that the original expression for $\Omega(\f)$ in Eq.~\eqref{TheIntegral} is analytic at $\f = 0$.

\section{Generalization of the result for complex chemical potential $\tilde{\mu}$}

Given the result of Eq.~\eqref{TheResult}, we can now easily generalize our calculation for $\mu \neq 0$. The required integral takes the form
\begin{equation}
\Omega(\f,\mu) = \frac{1}{\beta} \int{\frac{\d^3 p}{(2\pi)^3} \ln\left(1-\e^{\i \f} \e^{- \beta (E - \mu)}\right)} + \{\f \to - \f, \mu \to -\mu\}.
\label{eq:NonZero}
\end{equation}
Here, it should be noted that the (real) chemical potential of antiparticles differs in sign with respect to that of particles. In other words, we have
\begin{equation}
\Omega(\f,\mu) = \Omega(\f)\big|_{\f \to \f - \i \beta \mu},
\label{eq:TheVeza}
\end{equation}
since $\Omega(\f)$ is an analytic function in $\f$. Therefore, we arrive at our final result
\begin{align}
\Omega(\f,\mu) \equiv \Omega(\tilde{\mu}) &= \frac{2\pi^2}{3\beta^4}B_4\left(\frac{-\i \beta \tilde{\mu}}{2\pi}\right) + \frac{1}{2\beta^2}B_2\left(\frac{- \i \beta \tilde{\mu}}{2\pi}\right)m^2 + \frac{1}{32\pi^2}\left(\frac{3}{2}+2\ln 4\pi - \ln \beta^2 m^2 \right)m^4 \nonumber \\
&+ \frac{m^4}{16 \pi^2} \sum_{k=0}^\infty \frac{(-1)^k}{k!(k+2)!} \left[\psi^{(2k)}\left(\frac{- \i \beta \tilde{\mu}}{2\pi}\right) + \psi^{(2k)}\left(1+\frac{\i \beta \tilde{\mu}}{2\pi}\right)\right]\left(\frac{\beta m}{4\pi}\right)^{2k},
\label{FinalResult}
\end{align}
where $\tilde{\mu} = \mu + \i \f/\beta$ is the complex chemical potential. Equation \eqref{FinalResult} represents the main result of this work. The radius of convergence of the series in Eq.~\eqref{FinalResult} is finite and dependent on the value of the parameter $\beta \tilde{\mu}$. As a check, we consider the case of massless ($m=0$) fermions ($\f = \pi$) with non-zero chemical potential $\mu$, and find that the result agrees with the case considered in Ref.~\cite{Benic:2012ec}
\begin{equation}
\Omega(\pi,\mu)\big|_{m=0} = \frac{2\pi^2}{3\beta^4}B_4\left(\frac{\pi - \i \beta \mu}{2\pi}\right) = \frac{7\pi^2}{360 \beta^4} + \frac{\mu^2}{12\beta^2} + \frac{\mu^4}{24\pi^2}.
\label{eq:Khm}
\end{equation} 

\section*{Discussion and outlook}

In this paper, we have obtained an exact high temperature expansion for a one-loop thermodynamic potential $\Omega(\tilde{\mu})$ with complex chemical potential $\tilde{\mu}$. The final expression for $\Omega(\tilde{\mu})$ is given in Eq.~\eqref{FinalResult} and is the main result of this work. It is for the first time that the generalized thermodynamic potential is given as a single compact sum the coefficients of which are analytical functions of $\tilde{\mu}$, consisting of polynomials and polygamma fucntions, decoupled from the physical expansion parameter $\beta m$. This is what makes our solution convenient for the analysis of phase transitions for arbitrary $\tilde{\mu}$. We have used this fact to investigate the origin of the $m^3$ term in the boson case. For this, it was crucial to have an exact expansion of the type given in Eq.~\eqref{TheForm} in all orders of $\beta m$. Furthermore, the analytic nature of the coefficients $a_k(\f)$ allowed us to perform analytical continuation from purely imaginary to complex chemical potential almost effortlessly.

Earlier approaches have led to more complicated and less transparent results. The author of Ref.~\cite{Actor:1987cf} gave an expression for $\Omega(\tilde{\mu})$ in terms of triple expansion over $\beta m$, $\f$ and $\mu$. Our result can, therefore, be considered as a resummation of the two expansions, namely over $\f$ and $\mu$ into a compact function. Furthermore, in case of zero $\mu$, a result \cite{Meisinger:2001fi} gave a single sum representation of $\Omega(\f)$ in which the physical variable $\beta m$ was entangled with the parameter $\f$ in a nontrivial way. Moreover, the terms in the expansion were not analytic functions and the analytic continuation to nonzero $\mu$ could not be obtained so easily as in our case. Therefore, we are of the opinion that our result represents a significant improvement over earlier results and fills the gap in the literature concerning the compact analytic expression for generalized thermodynamic potential.

It would certainly be interesting to extend our calculation to an arbitrary number of spatial dimensions $d$. In the most interesting case $d=2$, the parameter $\f$ might be related to the anyon statistics phase and need not be interpreted as an imaginary chemical potential, as we have implicitly done in the paper. This leaves the possibility that the analysis performed in this paper could be relevant to the theory of phase transitions in $(2+1)$ dimensions, especially in the light of the famous Mermin--Wagner theorem \cite{PhysRevLett.17.1133,PhysRev.158.383,Coleman} which states that there are no phase transitions for boson and fermion systems in $d \leq 2$. We leave, however, this considerations for some future work.

\begin{acknowledgments}
The author would like to thank Sanjin Beni\'c for introducing him to the problem and discussing its physical applications at length. Also, the author has benefited greatly from many useful discussions with Goran Duplan\v ci\' c, Tajron Juri\'c, Matko Milin, Bene Ni\v zi\' c and Petar \v Zugec. Valuable suggestions and comments were given by an anonymous referee. This work was supported by the Croatian Ministry of Science, Education and Sport under the contract no.~098-0982390-2864.
\end{acknowledgments}
\appendix*
\section{Evaluating the second sum in Eq.~\eqref{TheLimit}}
The second sum on the r.h.s.~in Eq.~\eqref{TheLimit} can be evaluated as follows. Putting $x = \left(\frac{\beta m}{2 \f}\right)^2$, we have a sequence of equalities
\begin{align}
\mathcal{S}(x) &= \sum_{k=0}^\infty \frac{1}{k!} \frac{(2k)!}{(k+2)!} (-x)^k = \sum_{k=0}^\infty \frac{(2k-1)!!}{(k+2)!} (-2x)^k \nonumber \\
&= \frac{1}{4x^2}\sum_{k=0}^\infty \frac{(2k-5)!!}{k!} (-2x)^k - \frac{(-5)!!}{4x^2} + \frac{(-3)!!}{2x} \nonumber \\
&= \frac{1}{32 x^2} \frac{1}{\Gamma \left(\frac{3}{2}\right)} \sum_{k=0}^\infty \frac{\Gamma\left(k-\frac{3}{2}\right)}{k!} (-4x)^k - \frac{1}{12 x^2} - \frac{1}{2x}\nonumber \\
&= \frac{\pi}{32 x^2} \frac{1}{\Gamma \left(\frac{3}{2}\right) \Gamma \left(\frac{5}{2}\right)} \sum_{k=0}^\infty \frac{\Gamma\left(\frac{5}{2}\right)}{\Gamma(k+1) \Gamma\left(\frac{5}{2}-k\right)} \frac{(-4x)^k}{\sin\left(\pi k - \frac{3}{2}\pi\right)} - \frac{1}{12 x^2} - \frac{1}{2x} \nonumber \\
& = \frac{\pi}{32 x^2} \frac{1}{\Gamma \left(\frac{3}{2}\right) \Gamma \left(\frac{5}{2}\right)} \sum_{k=0}^\infty \binom{3/2}{k} (4x)^k - \frac{1}{12 x^2} - \frac{1}{2x} \nonumber \\
& = \frac{1}{12 x^2} \left(1+4x\right)^{3/2} - \frac{1}{12 x^2} - \frac{1}{2x}.
\end{align}
Here we have used the definition of the double factorial, its relation to the gamma function and the reflection formula for the gamma function. For the case $\f \ll 1$, i.e. for $x \gg 1$, we find
\begin{equation}
\mathcal{S}(x) \approx \frac{2}{3\sqrt{x}},
\end{equation}
which then leads to the Eq.~\eqref{Approx}.

It should be noted that the radius of convergence of the above sum is $R = \frac{1}{4}$, making the sum formally divergent for $x \gg 1$. This is a consequence of taking the termwise limit of Eq.~\eqref{TheResult}, which is not allowed since the series is not uniformly convergent. However, the naive summation employed here is also justified by the arguments given in Section I.B.

\bibliography{Integralcina}
\end{document}